\documentclass[aps,pra,superscriptaddress,amsmath,amssymb,preprintnumbers,twocolumn]{revtex4}
\usepackage{amssymb}
\usepackage{booktabs}
\usepackage{graphicx}
\usepackage{bm}
\usepackage{color}
\usepackage{cancel}
\usepackage{slashed}
\usepackage{subfigure}

\newcommand{\Ignore}[1]{}

\newcommand{\Ket}[1]{\left\vert #1\right\rangle}

\newcommand{\ii}{\mathrm{i}}
\newcommand{\ee}{\mathrm{e}}

\begin{document}

\title{Degenerate Landau-Zener model in the presence of quantum noise}

\author{Benedetto Militello}
\address{Universit\`a degli Studi di Palermo, Dipartimento di Fisica e Chimica - Emilio Segr\`e, Via Archirafi 36, 90123 Palermo, Italia}
\address{I.N.F.N. Sezione di Catania, Via Santa Sofia 64, I-95123 Catania, Italia}

\begin{abstract}
The degenerate Landau-Zener-Majorana-St\"uckelberg model consists of two degenerate energy levels whose energies vary with time and in the presence of an interaction which couples the states of the two levels. In the adiabatic limit, it allows for the populations transfer from states of one level to states of the other level. The presence of an interaction with the environment influences the efficiency of the process. Nevertheless, identification of possible decoherence-free subspaces permits to engineer coupling schemes for which quantum noise can be minimized.
\end{abstract}

\maketitle

\section{Introduction}\label{sec:introduction}

A two-state quantum system with time-dependent energies and subjected to an interaction which induces transitions between the two states undergoes an evolution which can essentially consist in the adiabatic following of the eigenstates or involve diabatic transitions between them, depending on the chirping rate of the energies. The amount of such passage of population through diabatic processes can be evaluated through the very famous formula independently found by Landau, Zener, Majorana and St\"uckelberg (LZMS) in the same year~\cite{ref:Landau,ref:Zener,ref:Majo,ref:Stuck}. In the original problem, the diabatic energies were assumed to be changing linearly with time, the coupling strength between the two states was supposed to be constant and the time was hypothesized to span a very large interval, virtually ranging from $-\infty$ to $+\infty$. From there on, several assumptions have been relaxed, leading to many different generalization of the original LZMS model.  

Nonlinear time-dependence of the diabatic energies have been considered~\cite{ref:Vitanov1999b} as well as finiteness of the time interval associated to the experiment~\cite{ref:Vitanov1996,ref:Vitanov1999a}. Models where the system undergoing an energy crossing is governed by nonlinear equations have been analyzed~\cite{ref:Ishkhanyan2004,ref:MilitelloPRE2018}. Non-Hermitian Hamiltonian models have been considered too~\cite{ref:Toro2017}.
The LZMS model describes an avoided crossing, since on the one hand the diabatic (bare) energies cross while the adiabatic (dressed) energies do not. Variants have analyzed, consisting in the hidden crossing model, where neither the diabatic nor the adiabatic energies cross~\cite{ref:Fishman1990,ref:Bouwm1995}, and in the total crossing model, where both the diabatic and the adiabatic energies cross~\cite{ref:Militello2015a}. The latter model requires that the coupling strength (the off-diagonal term of the $2\times2$ Hamiltonian) is time-dependent and vanishes at the same time with the diagonal terms.
Extended models where more than two quantum states are assumed to be involved in the evolution have been introduced. Majorana studied the dynamics of a spin-$j$ immersed in a magnetic field in such a situation where a multi-state avoided crossing occurs~\cite{ref:Majo}. Multi-state systems undergoing a series of pairwise crossings, hence allowing for the so called independent crossing approximation, have been considered~\cite{ref:Demkov1967,ref:Brundobler1993}. Moreover, proper multi-state avoided crossings have been studied in details, under special hypotheses about the coupling scheme. In particular, the $N$-state bow-tie model, where one state is coupled to the remaining $N-1$ (or two states are coupled to the remaining $N-2$) which do not couple to each other, have been extensively investigated~\cite{ref:Carroll1986a,ref:Carroll1986b,ref:Demkov2000}.
Several other specific schemes involving many levels have been proposed and studied in details~\cite{ref:Shytov2004,ref:Sin2015,ref:Li2017,ref:Sin2017}, as well as effective LZMS models able to describe the dynamics of spin-boson systems governed by the time-dependent Rabi Hamiltonian~\cite{ref:Dodo2016} or Tavis-Cummings model~\cite{ref:Sun2016,ref:Sin2016}.
The interest in the LZMS model is witnessed by several experiments that have been developed with systems which are adiabatically or quasi-adiabatically driven in the proximity of avoided crossings~\cite{ref:Berns2008,ref:Song2016,ref:Zhang2017}.

Over the decades, several papers have investigated the role of quantum noise on adiabatic evolutions in general~\cite{ref:Lidar,ref:Florio,ref:MilitelloPRA2010,ref:ScalaOpts2011,ref:MilitelloPScr2011,ref:Wild2016} and in the specific case of the two-state LZMS processes~\cite{ref:Ao1991,ref:Potro2007,ref:Wubs2006,ref:Saito2007,ref:Lacour2007,ref:Nel2009,ref:ScalaPRA2011}. Recently, some contributions have appeared on the effects of the interaction with the environment for multi-state LZMS processes, in different configurations~\cite{ref:Ashhab2016,ref:Militello2019c} and exploiting non-Hermitian Hamiltonian models~\cite{ref:Militello2019a,ref:Militello2019b}. 

A different and intriguing scenario is that addressed as the degenerate Landau-Zener model. It is realized when two degenerate levels cross, provided some interaction is present which couples states of one level to states of the other, but does not couple states belonging to the same level~\cite{ref:Vasilev2007}. This model allows for mapping linear combinations of the states of one level to linear combinations of the other.

In this paper we study the degenerate Landau-Zener model in the presence of quantum noise. Depending on the structure of the system-environment interaction we are able to identify special states which are not subjected (or are marginally subjected) to the noise. In practice, we single out the presence of decoherence-free subspaces. Suitable coupling schemes realized through coherent fields allow for exploiting such protected zones of the relevant Hilbert space to obtain efficient population transfer. In sec.~\ref{sec:model} we introduce the dissipative degenerate Landau-Zener model. In section \ref{sec:2x2} we discuss a specific case, focusing on the configuration where a twofold degenerate level is coupled to another twofold degenerate one. We provide the explicit transformations that diagonalize the Hamiltonian of the system and the system-environment interaction term, and then identify possible decoherence-free subspaces. The relevant dynamics are evaluated numerically in order to compare them with the theoretical predictions. Finally, in sec.~\ref{sec:conclusions} we discuss the results and the possibility to explot them in order to obtain information about the system-environment couplig.

\section{The Model}\label{sec:model}

\begin{figure}[h]
\includegraphics[width=0.45\textwidth, angle=0]{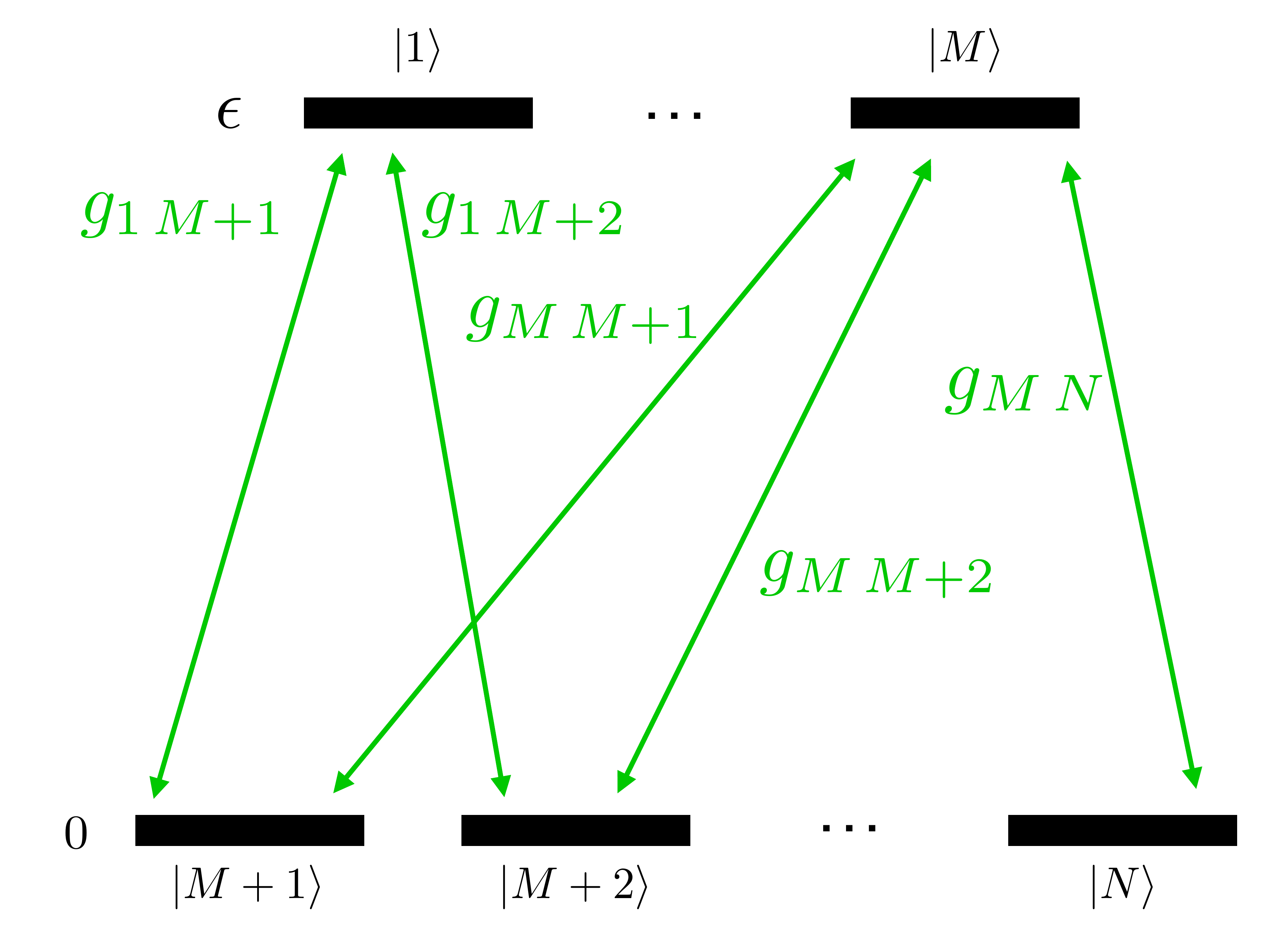}
\caption{(Color online) Scheme for coherent couplings: the states of a degenerate eigenspace are coupled to the states of a degenerate eigenspace with lower energy; the relevant coupling constants are $g_{ij}$. No coupling between states of the same subspace is considered.
}\label{fig:GenScheme}
\end{figure}

{\it Idela case --- } The general $M:(N-M)$ model refer to a $N$-state system, with two degenerate subspaces of degeneracies $M$ and $N-M$, subjected to external fields which produce couplings between the states of the two subspaces. The external fields never couple two states belonging to the same subspace, and therefore the relevant Hamiltonian has the form:
\begin{equation}
\hat{H} = \left(
\begin{array}{cc}
\epsilon \, \mathbf{I}_{M} & \mathbf{G} \\
\mathbf{G}^\dag & \mathbf{0}_{N-M}
\end{array} \right)\,,
\end{equation}
where $\mathbf{G}$ is a $M\times(N-M)$ matrix and where the two bare energies of the two levels are assumed to be $\epsilon$ and $0$. The operators $\mathbf{I}_M$ and $\mathbf{0}_{N-M}$ are the identity of an $M$-dimensional space and the ($N-M$)-dimensional null operator, respectively. 
In Fig.~\ref{fig:GenScheme} is represented an example of coupling scheme. 
This sort of interaction pattern has been studied in the past. Morris and Shore~\cite{ref:Morris1982} have developed an analytical treatment based on a suitable transformation that diagonalizes the matrix $\mathbf{G}$. A generalization to the multi-level state has also been presented~\cite{ref:Rangelov2006} . On this basis, Vasilev {\it et al}~\cite{ref:Vasilev2007} have used this scheme assuming a time-dependent $\epsilon=\kappa t$, leading to the degenerate Landau-Zener model, which is essentially equivalent to a series of independent two-state Landau-Zener models between states of the two levels. Indeed, following the Morris and Shore theory, once the matrices $\mathbf{G}^\dag \mathbf{G}$ and $\mathbf{G}\mathbf{G}^\dag$ are diagonalized, we can find out the transformation that can put $\mathbf{G}$ in a diagonal form. At this point, it is easy to see that when $t$ spans a large time interval $[-t_0, t_0]$ ($\kappa t_0 / \Omega \gg 1$) with a sufficiently small chirping rate ($\kappa/\Omega^2 \ll 1$), then states of the upper level are adiabatically mapped into states of the lower level and vice versa.

{\it Quantum Noise --- } We will assume that the system is interacting with its environment in a way similar to that of the coherent fields, i.e., connecting only states of one level to states of the other level. Therefore, the system-environment Hamiltonian can be assumed having the form 
\begin{equation}
H_I = \lambda \hat{X} \otimes \hat{B} \,, 
\end{equation}
with
\begin{equation}
\hat{X} = \left(
\begin{array}{cc}
\mathbf{0}_{M} & \mathbf{W} \\
\mathbf{W}^\dag & \mathbf{0}_{N-M}
\end{array} \right)\,,
\end{equation}
where $\mathbf{W}$ is a $M\times(N-M)$ matrix. The environment is modeled as an infinite set of bosonic modes, as usual for atomic or pseudo-atomic systems interacting with the electromagnetic field.

The non-unitary evolution of the system can be straightforwardly evaluated through the Davies and Spohn theory~\cite{ref:DaviesSpohn}. According to such theory, assuming that the typical environment correlation time is very much smaller than the time scale of the Hamiltonian modification, the system can be thought of as frozen with respect to the bath. The relevant master equation can then be obtained with the standard approach~\cite{ref:Petru,ref:Gardiner}, where the jump operators and the decay rates are time-dependent and related to the instantaneous eigenvalues and eigenstates of the system Hamiltonian.
This approach has been extensively used in the study of quantum gates~\cite{ref:Florio}, Landau-Zener processes~\cite{ref:ScalaPRA2011,ref:Militello2019c} and STIRAP manipulations~\cite{ref:MilitelloPRA2010,ref:ScalaOpts2011,ref:MilitelloPScr2011}. The relevant master equation has the following form:
\begin{subequations}
\begin{eqnarray}
\dot\rho &=& -\ii [\hat{H}, \rho] + \sum_{i\not=j} \gamma_{ij} {\cal D}(\hat{X}_{ij}, \rho) \,,
\end{eqnarray}
where
\begin{eqnarray}
{\cal D}(\hat{O}, \rho) &=& \hat{O} \rho \hat{O}^\dag - \frac{1}{2} \{ \hat{O}^\dag \hat{O}, \rho \} \,. \\
\hat{X}_{ij} &=& \hat{\Pi}_i \, \hat{X} \, \hat{\Pi}_j \,, \\
 \hat{\Pi}_i \, \hat{H} &=& \hat{H} \, \hat{\Pi}_i = \epsilon_i\hat{\Pi}_i\,,
\end{eqnarray}
and
\begin{eqnarray}
&& \gamma_{ij} = \int_{-\infty}^\infty \ee^{\ii\omega_{ij}s} \mathrm{tr}_E [\hat{B}(s)\hat{B}(0) \rho_E] \mathrm{d}s \,,
\end{eqnarray}
\end{subequations}
where $T$ is the bath temperature, $\rho_E$ is the thermal state of the environment, $\omega_{ij}=\omega_j-\omega_i$ is a generic transition frequency of the system and $\hat{B}(s)$ is the operator $\hat{B}$ in the interaction picture at time $s$. It turns out $\gamma_{ij} = \lambda^2 \, |\alpha(\omega_{ij})|^2 D(\omega_{ij}) N(\omega_{ij}, T)$, where $\alpha(\omega_{ij}$ is the coupling strength of the system with the modes of frequency $\omega_{ij}$, $D(\omega_{ij})$ is the density of modes at frequency $\omega_{ij}$ and $N(\omega_{ij}, T)=\mathrm{sign}{\omega_{ij}}/[1-\ee^{-\omega_{ij}/(k_B T)}]$ is the average number of bosons at frequency $\omega_{ij}$. To be consistent with the hypothesis of short correlation time (necessary for the Davies and Spohn theory), which is related to the hypothesis of flat spectrum, we will assume that the quantity $|\alpha(\omega_{ij})|^2 D(\omega_{ij})$ does not depend on $\omega_{ij}$. Therefore, we will be in a condition to introduce the parameter $\gamma\equiv  \lambda^2 \, |\alpha(\omega_{ij})|^2 D(\omega_{ij}) $.

{\it Noise-free subspace --- } Sometimes, it is possible to identify possible subspaces of the two-level system not sensitive to the interaction with the environment, then finding out decoherence-free~\cite{ref:Lidar1998,ref:Lu1997} or system-environment interaction-free subspaces~\cite{ref:Chrusc2015}. To this purpose, it is convenient to put the matrix $\mathbf{W}$ of $\hat{X}$ in diagonal form, to find out which states of the system are immune to the noise, if any. Once such possible decoherence-free states are identified, we can choose the coherent couplings responsible for the interaction terms in $\hat{H}$ in such a way to connect the stable states. This will guarantee a noise-free population transfer. Of course, there are situations where no decoherence-free subspace is present.

\section{The $2:2$ case}\label{sec:2x2}

\begin{figure}[h]
\includegraphics[width=0.45\textwidth, angle=0]{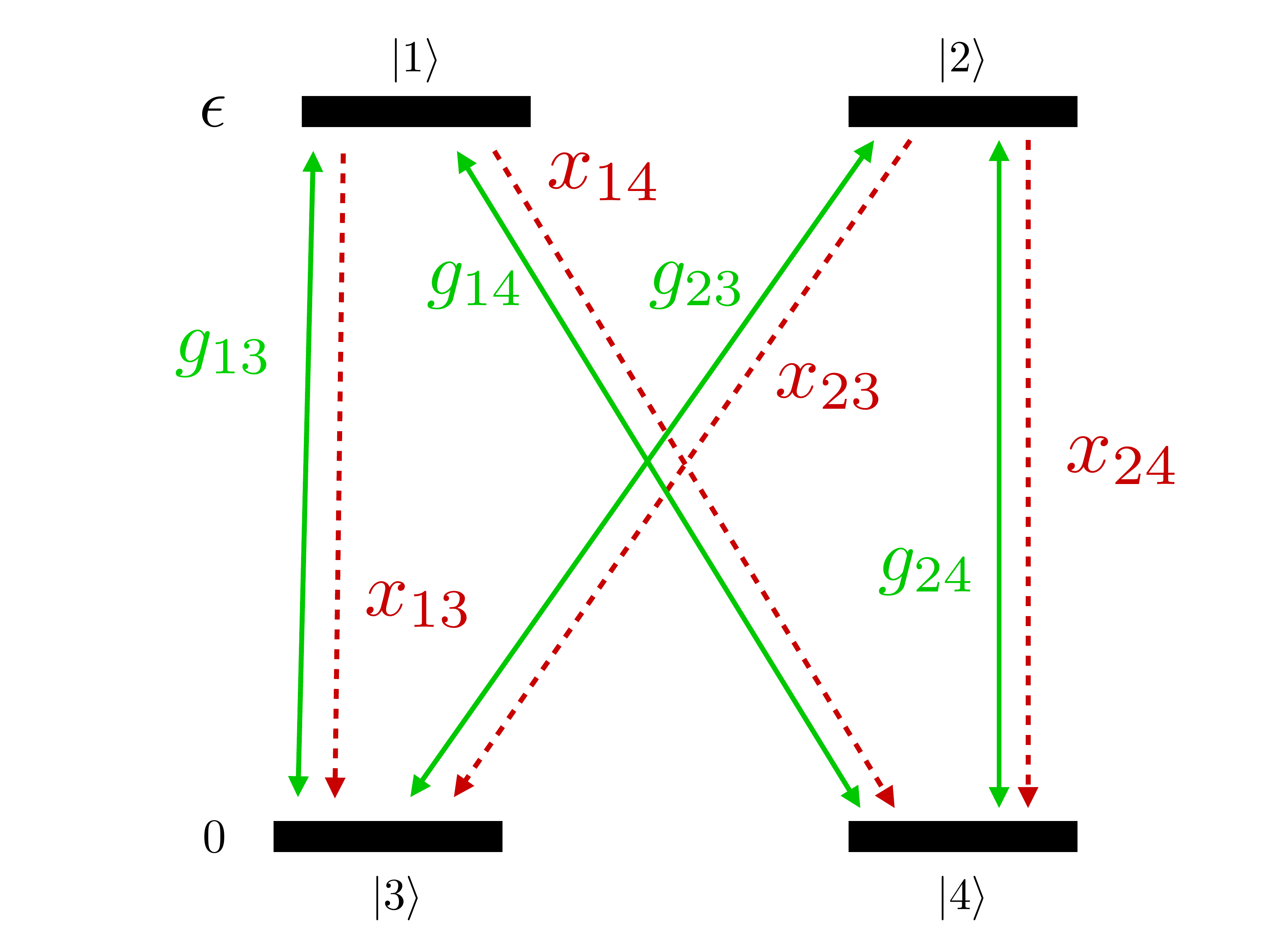}
\caption{(Color online) Coupling scheme for two levels with twofold degeneration. The green solid rows represent the coherent couplings, while the red dashed rows represent the environment-induced interactions. According to the general scheme, no coupling between states of the same level is considered.
}\label{fig:2x2Scheme}
\end{figure}

The first case we will focus on corresponds to $N=4$ and $M=2$ and to the coupling scheme in  Fig.~\ref{fig:2x2Scheme}. The relevant Hamiltonian and interaction with the environment are:
\begin{equation}
\hat{H} = \left(
\begin{array}{cccc}
\epsilon & 0 & g_{13} & g_{14} \\ 
0 & \epsilon & g_{23} & g_{24} \\ 
g_{13}^* & g_{23}^* & 0 & 0 \\ 
g_{14}^* & g_{24}^* & 0 & 0  
\end{array} \right)\,,
\end{equation}
and
\begin{equation}
\hat{X} = \left(
\begin{array}{cccc}
\epsilon & 0 & x_{13} & x_{14}\\ 
0 & \epsilon & x_{23} & x_{24} \\ 
x_{13} & x_{23} & 0 & 0 \\ 
x_{14} & x_{24} & 0 & 0  
\end{array} \right)\,.
\end{equation}
For the sake of simplicity, we will assume all that $x_{ij}$ and $w_{ij}$ are real. To put this operator in the form where each of two orthogonal states in the subspace $\{\Ket{1},\Ket{2}\}$ is coupled to one of two orthogonal states in the subspace $\{ \Ket{3}, \Ket{4} \}$, we apply a unitary transformation of the form
\begin{eqnarray}
\hat{R}(\xi,\chi) = 
\left(
\begin{array}{cccc}
\cos\xi & \sin\xi & 0 & 0 \\ 
-\sin\xi & \cos\xi & 0 & 0 \\ 
0 & 0 & \cos\chi & \sin\chi \\ 
0 & 0 & -\sin\chi & \cos\chi  
\end{array} \right)\,, 
\end{eqnarray}
which leads to 
\begin{subequations}
\begin{equation}
\tilde{X} = \hat{R}(\xi,\chi) \hat{X} \hat{R}^{-1}(\xi,\chi) = \left(
\begin{array}{cccc}
\epsilon & 0 & \tilde{x}_{13} & \tilde{x}_{14} \\ 
0 & \epsilon & \tilde{x}_{23} & \tilde{x}_{24} \\ 
\tilde{x}_{13}^* & \tilde{x}_{23}^* & 0 & 0 \\ 
\tilde{x}_{14}^* & \tilde{x}_{24}^* & 0 & 0  
\end{array} \right)\,,
\end{equation}
with
\begin{eqnarray}
\nonumber
\tilde{x}_{13} &=& \cos\chi (x_{13} \cos\xi +  x_{23} \sin\xi) \\
&+& \sin\chi (x_{14} \cos\xi +  x_{24} \sin\xi) \, ,  \\
\nonumber
\tilde{x}_{14} &=& \cos\chi (x_{14} \cos\xi +  x_{24} \sin\xi) \\
&-& \sin\chi (x_{13} \cos\xi +  x_{23} \sin\xi) \, , \\
\nonumber
\tilde{x}_{23} &=& \cos\chi (x_{23} \cos\xi -  x_{13} \sin\xi) \\
&+& \sin\chi (x_{24} \cos\xi -  x_{14} \sin\xi) \, , \\
\nonumber
\tilde{x}_{24} &=& \cos\chi (x_{24} \cos\xi - x_{14} \sin\xi) \\
&-& \sin\chi (x_{23} \cos\xi -  x_{13} \sin\xi) \, . 
\end{eqnarray}
\end{subequations}

Now, we impose $\tilde{x}_{14}=\tilde{x}_{23}=0$, which is obtained when
\begin{subequations}
\begin{eqnarray}
\tan\xi &=& \frac{A+B-\sqrt{C}}{2D} \, , \\
\tan\chi &=& \frac{A-B-\sqrt{C}}{2E}  \, ,
\end{eqnarray}
with
\begin{eqnarray}
A &=& x_{24}^2 - x_{13}^2 \, , \\
B &=& x_{23}^2 - x_{14}^2 \, , \\
\nonumber
C &=& \left[(x_{14} + x_{23})^2 + (x_{13} - x_{24})^2\right] \\
   &\times& \left[(x_{14} - x_{23})^2 + (x_{13} + x_{24})^2\right] \, , \\
D &=& x_{13} x_{23} + x_{14} x_{24}  \, , \\
E &=& x_{13} x_{14} + x_{23} x_{24}  \, .
\end{eqnarray}
\end{subequations}

This solution is allowed provided $x_{13} x_{23} + x_{14} x_{24} \not=0$ and $x_{13} x_{14} + x_{23} x_{24}\not=0$\,. If these conditions are not satisfied, a direct solution can be easily found. In fact, if $x_{13} x_{23} + x_{14} x_{24} =0$ it means that the states $(x_{13}\Ket{3}+x_{14}\Ket{4})/\sqrt{x_{13}^2+x_{14}^2}$ and $(x_{23}\Ket{3}+x_{24}\Ket{4})/\sqrt{x_{23}^2+x_{24}^2}$ are orthogonal, and such two states are coupled to $\Ket{1}$ and $\Ket{2}$, respectively. On the other hand, if $x_{13} x_{14} + x_{23} x_{24}=0$ then the states $(x_{13}\Ket{1}+x_{23}\Ket{2})/\sqrt{x_{13}^2+x_{23}^2}$ and $(x_{14}\Ket{1}+x_{24}\Ket{2})/\sqrt{x_{14}^2+x_{24}^2}$ are orthogonal and coupled to the orthogonal states $\Ket{3}$ and $\Ket{4}$, respectively. 

Let us consider the special case where $x_{ij}=1/2$, $\forall i,j$, which implies $\xi=\chi=-\pi/4$ and then $\tilde{x}_{13}=\tilde{x}_{14}=\tilde{x}_{23}=0$, $\tilde{x}_{24}=1$. 
Therefore, in the rotated picture, the environment does not affect the states $\Ket{1}$ and $\Ket{3}$, and an Hamiltonian $\tilde{H}$ characterized by $\tilde{g}_{13}=g\not=0$ and all the other $\tilde{g}_{ij}$ equal to zero is responsible for a noise-free population transfer. This corresponds to $\hat{H} = \hat{R}(\pi/4,\pi/4) \tilde{H} \hat{R}^{-1}(\pi/4,\pi/4)$, having the following coupling constants: $g_{13}=g_{24}=g/2$, and $g_{14}=g_{23}=-g/2$. Such Hamiltonian carries population from $(\Ket{1}-\Ket{2})/\sqrt{2}$ to $(\Ket{3}-\Ket{4})/\sqrt{2}$, and vice versa, without environmental effects.

\begin{figure}[h]
\subfigure[]{\includegraphics[width=0.45\textwidth, angle=0]{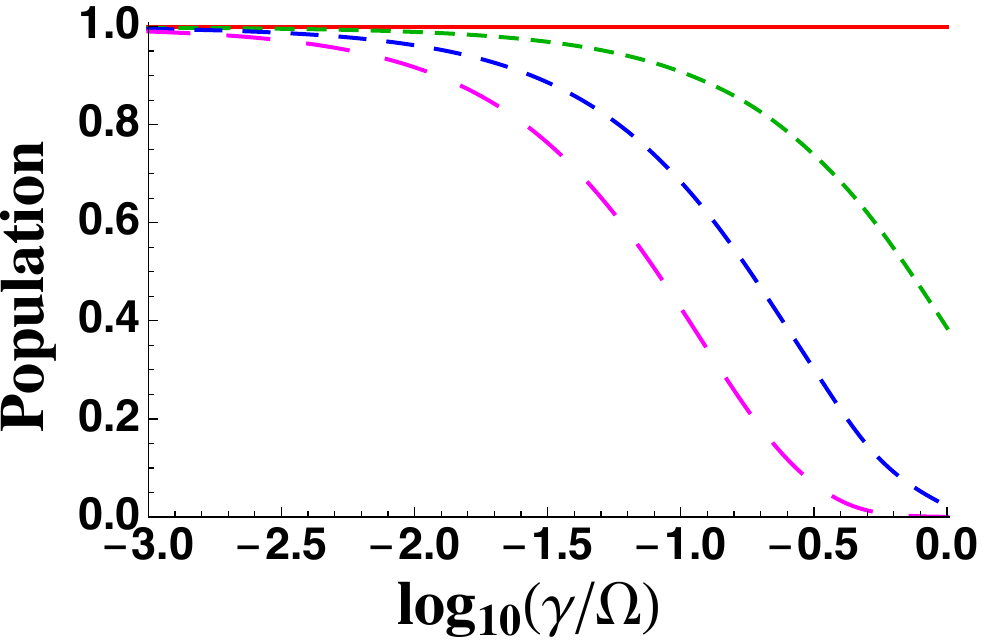}}
\subfigure[]{\includegraphics[width=0.45\textwidth, angle=0]{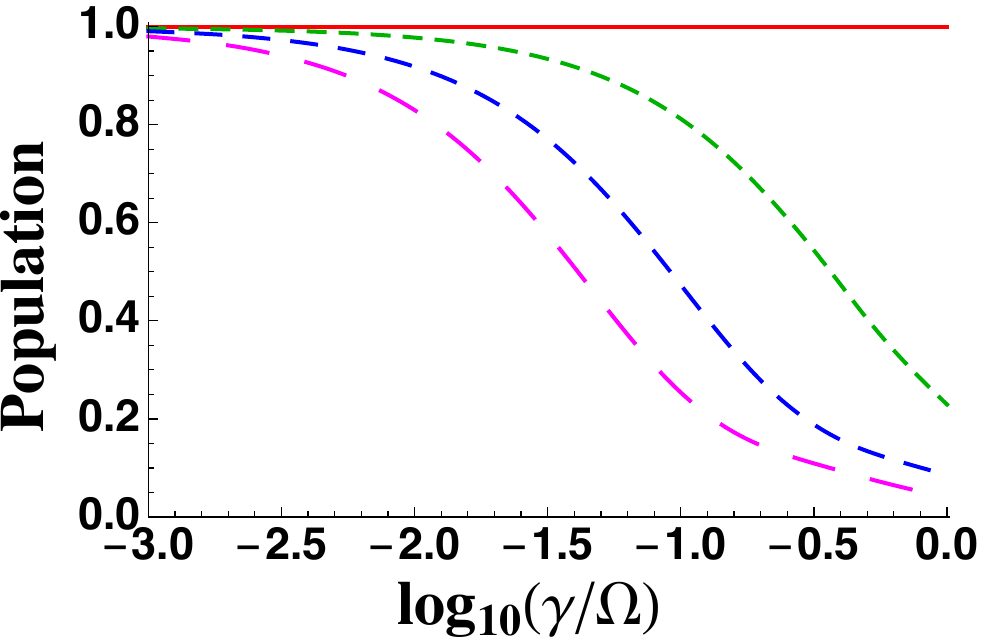}}
\caption{(Color online) Population of the target state $\Ket{\psi_f}=-\sin\xi_g\Ket{1} + \cos\xi_g\Ket{2}$ when the system starts in $\Ket{\psi_i}=-\sin\chi_g\Ket{3} + \cos\chi_g\Ket{4}$ as a function of $\gamma$ (in units of $\Omega$ and in logarithmic scale), for the coupling scheme characterized by $x_{ij}=1/2$, $g_{13}=2^{-1/2} g \,  \cos(\pi/4+\delta_1)$, $g_{14}=2^{-1/2} g \,  \sin(\pi/4+\delta_1)$, $g_{23}=2^{-1/2} g \,  \cos(3\pi/4+\delta_2)$ and $g_{24}=2^{-1/2} g \, \sin(3\pi/4+\delta_2)$. Different values of the parameters $\delta_1$ and $\delta_2$ are considered: $\delta_1=\delta_2=0$ (solid red line), $\delta_1=\pi/36$ and $\delta_2=0$ (dotted green line), $\delta_1=\pi/18$ and $\delta_2=0$ (dashed blue line), and $\delta_1=\pi/36$ and $\delta_2=\pi/18$ (long dashed pink line). Different values of temperature have been attributed to the environment: $k_B T/\Omega = 0.001$ (a) and $k_B T / \Omega = 10$ (b). The other parameters are $\kappa/\Omega^2=0.1$ and $\kappa t_0/\Omega = 50$.
}\label{fig:Dissipative2x2-H}
\end{figure}

It is interesting to study the robustness is the process with respect to imperfections, i.e., what happens if the coupling constants are not exactly in the required relation to exploit the presence of the decoherence-free subspace. In particular, assume $g_{13}=2^{-1/2} g \,  \cos(\pi/4+\delta_1)$, $g_{14}=2^{-1/2} g \,  \sin(\pi/4+\delta_1)$, $g_{23}=2^{-1/2} g \,  \cos(3\pi/4+\delta_2)$ and $g_{24}=2^{-1/2} g \, \sin(3\pi/4+\delta_2)$. 
In Fig.~\ref{fig:Dissipative2x2-H} is represented the efficiency of the population transfer as a function of the decay rate $\gamma$, for  different values of $\delta_1$ and $\delta_2$, considering the initial state $\Ket{\psi_i} = -\sin\chi_g\Ket{3} + \cos\chi_g\Ket{4}$, where $\chi_g$ is the angle $\chi$ evaluated with respect to coefficients $g_{ij}$'s in place of $x_{ij}$'s. Consequently, the target state is $\Ket{\psi_i} = -\sin\xi_g\Ket{1} + \cos\xi_g\Ket{2}$, with $\xi_g$ evaluated with respect to coefficients $g_{ij}$. For $\delta_1=\delta_2=0$, the efficiency is always essentially $100\%$, irrespectively of the parameter $\gamma$, while for $\chi\not=-\pi/4$, i.e., $\delta_1\not=0$ and/or $\delta_2\not=-\pi/4$, the efficiency reduces as $\gamma$ increases. In particular, higher values of $\delta_1$ or $\delta_2$ (implying a more significant distance from the optimal situation) the population transfer becomes more and more sensitive to the quantum noise. The behavior is quite similar for essentially zero temperature ($k_B T/\Omega = 0.001$ in Fig.~\ref{fig:Dissipative2x2-H}a) and for moderately high temperature  ($k_B T/\Omega = 10$ in Fig.~\ref{fig:Dissipative2x2-H}b).

\begin{figure}[h]
\subfigure[]{\includegraphics[width=0.45\textwidth, angle=0]{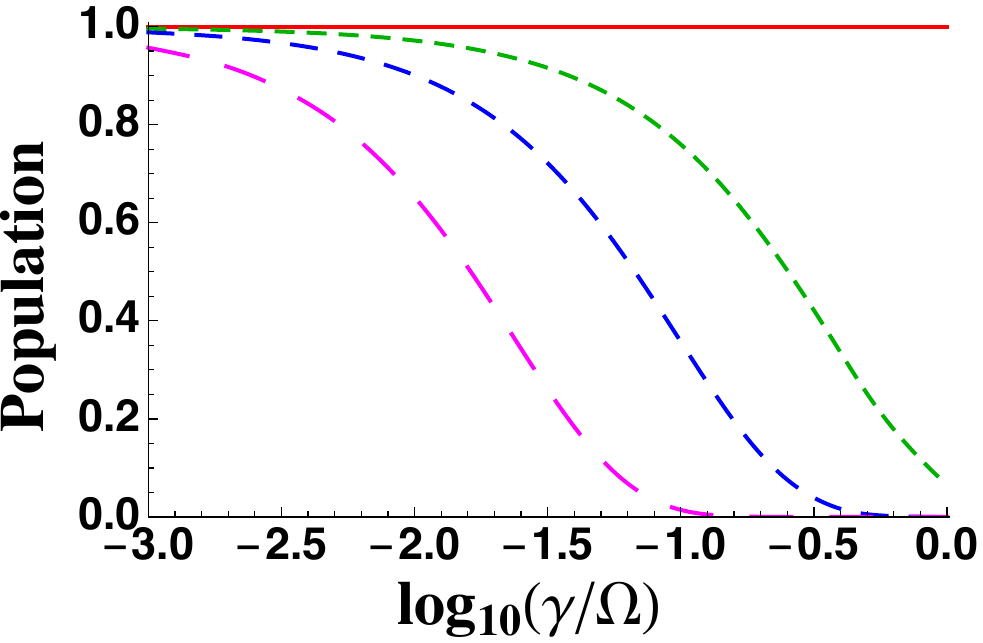}}
\caption{(Color online) Population of the target state $\Ket{\psi_f}=-\sin\xi_g\Ket{1} + \cos\xi_g\Ket{2}$ when the system starts in $\Ket{\psi_i}=-\sin\chi_g\Ket{3} + \cos\chi_g\Ket{4}$ as a function of $\gamma$ (in units of $\Omega$ and in logarithmic scale), for the coupling scheme characterized by $g_{13}=g_{24}=1/2$, $g_{14}=g_{23}=-1/2$ and $x_{13}=2^{-1/2} \cos(\pi/4+\delta_1)$, $x_{14}=2^{-1/2} \sin(\pi/4+\delta_1)$, $x_{23}=2^{-1/2} \cos(\pi/4+\delta_2)$, $x_{24}=2^{-1/2} \cos(\pi/4+\delta_2)$. Different values of the parameters $\delta_1$ and $\delta_2$ are considered: $\delta_1=\delta_2=0$ (solid red line), $\delta_1=\pi/36$ and $\delta_2=0$ (dotted green line), $\delta_1=\pi/36$ and $\delta_2=\pi/18$ (dashed blue line), and $\delta_1=0$ and $\delta_2=\pi/9$ (long dashed pink line). The other parameters are $\kappa/\Omega^2=0.1$, $\kappa t_0/\Omega = 50$ and $k_B T/\Omega = 0.001$.
}\label{fig:Dissipative2x2-X}
\end{figure}

We also consider the complementary situation where the coherent couplings perfectly satisfy the condition for having a noiseless population transfer under the assumption of $x_{ij}=1$ $\forall i, j$, but the real $\hat{X}$ operator slightly deviates from such a scheme.  In particular, we assume $g_{13}=g_{24}=1/2$, $g_{14}=g_{23}=-1/2$ and $x_{13}=2^{-1/2} \cos(\pi/4+\delta_1)$, $x_{14}=2^{-1/2} \sin(\pi/4+\delta_1)$, $x_{23}=2^{-1/2} \cos(\pi/4+\delta_2)$, $x_{24}=2^{-1/2} \cos(\pi/4+\delta_2)$. It is easy to check that in the general case (i.e., out of the case $\delta_1=\delta_2=0$) there is no decoherence-free subspace, since it turns out that both $\tilde{x}_{13}$ and $\tilde{x}_{24}$ are nonzero.
In Fig.~\ref{fig:Dissipative2x2-X} is plotted the efficiency as a function of $\gamma$ for different values of the parameters $\delta_1$ and $\delta_2$.  The initial state is $\Ket{\psi_i}=(\Ket{3}-\Ket{4})/\sqrt{2}$, while the target is $\Ket{\psi_f}=(\Ket{1}-\Ket{2})/\sqrt{2}$. It is well visible also in this case that, as soon as the conditions to have a noiseless process are not perfectly fulfilled, the system becomes more and more sensitive to the noise.

\section{Discussion}\label{sec:conclusions}

In this paper, we have analyzed the role of quantum noise in the degenerate Landau-Zener-Majorana-St\"uckelberg model consisting of two degenerate levels whose diabatic energies cross at some instant of time and in the presence of interactions that couple states of one level to states of the other level, never connecting states belonging to the same level. As expected, the interaction with the environment negatively affects the population transfer realizable through the adiabatic following of some eigenstate of the Hamiltonian when the diabatic energy of the upper level changes. Nevertheless, by knowing the structure of the system-environment interaction, one can identify possible noise-free regions of the system Hilbert space, which can allow to realize noiseless population transfer. We then focused on the special case of a twofold degenerate level interacting with another twofold degenerate one. In this specific scenario, we have found which coupling is optimal to exploit the presence of a possible noise-free subspace and how the population transfer is affected by deviations from such optimal interaction. Both essentially-zero and moderately-high temperatures have been considered, singling out quite similar behaviors.

It is also worth taking into account on another special scenario, which is the $1:(N-1)$ case, i.e., when the upper level is non degenerate while the lower is $(N-1)$-degenerate. The peculiarity of this situation is that the upper level (the singlet) is inevitably coupled to some state of the lower level, both coherently and through the environment. This leads to the impossibility of identifying a noise-free subspace for the population transfer from one level to the other. In spite of this, if one considers a process that starts in the upper level and is supposed to end up in some state of the lower level, at zero temperature the environment will push the system toward the states with lower energies and, in some cases, will help the coherent population transfer. Nevertheless, the opposite coherent process leading from the lower state to the upper one will be thwarted. This is the reason why in the $2:2$ case we preferred to consider population transfer from the lower to the upper level. 

We finally comment on the fact that our analysis can pave the way to a quantum noise unravelling, since by changing the coherent coupling scheme (i.e., the $g_{ij}$ parameters) and measuring the efficiency of the population transfer, it is possible to get information about the parameters of the system operator $\hat{X}$ involved in the system-environment interaction. In particular, experimentally finding decoherence-free subspaces gives specific constraints for the quantities $x_{ij}$.


\end{document}